\documentclass{iopart} 
\usepackage{epsfig}

\newcommand{\be}[1]{\begin{equation}\label{#1}}
\newcommand{\ee}{\end{equation}}  
\newcommand{\fig}[1]{fig.~\ref{#1}}
\newcommand{\tbl}[1]{table~\ref{#1}}
\newcommand{\bea}{\begin{eqnarray}}
\newcommand{\eea}{\end{eqnarray}}
\newcommand{\eq}[1]{(\ref{#1})}

\newcommand{\showlabel}[1]{}

\begin{document} 
\jl 2
\title{A quasi classical approach to electron impact ionization}
\author{Tiham\'{e}r Geyer\dag \ and Jan M Rost\ddag}
\address{\dag Weizmann Inst. of Science,
Dept. of Chemical Physics, Rehovot 76100, Israel}
\address{\ddag Max--Planck--Institute for the Physics of Complex Systems,
N\"othnitzer Str. 38, D--01187 Dresden, Germany}
\date{\LaTeX ed \today}
\begin{abstract}
\noindent A quasi classical approximation to quantum mechanical
scattering in the M\o{}ller formalism is developed. While keeping the
numerical advantage of a standard Classical--Trajectory--Monte--Carlo
calculation, our approach is no longer restricted to use stationary
initial distributions. This allows one to improve the results by using
better suited initial phase space distributions than the
microcanonical one and to gain insight into the collision mechanism by
studying the influence of different initial distributions on the cross
section. A comprehensive account of results for single, double and
triple differential cross sections for atomic hydrogen will be given,
in comparison with experiment and other theories.
\end{abstract}
\pacs{34.50, 34.80, 82.30}
%


\section{Introduction}

Electron impact ionization of hydrogen probes the scattering
properties of three charged particles without any influence from
passive core electrons or other perturbations. For this reason it has
served as a benchmark system to understand non--separable Coulomb
collision dynamics for a long time which is reflected in the enormous
amount of literature and the remarkable theoretical and experimental
success in handling these collisions. A milestone were the experiments
by Ehrhardt and his group obtaining fully differential cross sections
on an absolute scale \cite{EHR85, EHR86}. Theoretically, many
different approximate treatments have reached increasingly better
agreement with the experiment \cite{CAP84, BRA89, JON92, BER93, WHE93}
and fully numerical calculations, first available for the total cross
section \cite{BRA92}, have reached a break-through recently
\cite{RES99} for fully differential cross sections and are now
producing results for many impact energies and geometries \cite{BAE01a,
BAE01b, ISA01}.

Our goal with the formulation of a quasiclassical theory of electron
impact ionization is to open the way for scattering of more than three
particles as already done experimentally with double ionization by
electron impact \cite{LAH89, ULL97}. Furthermore, we aim at a
description which allows for a calculation of the entire scattering
information (i.e. the whole accessible parameter space in final angles
and energies) at once. This is motivated by similar experimental
capabilities made possible by the COLTRIMS technique which allows for
a new representation of scattering data (see e.g. \cite{DOR96,
DOR99}).

Clearly, with the present computational and theoretical tools this is
not possible for four our more particles. One has to fall back to
quasiclassical approximations based on the rather successful CTMC
method \cite{ABR66a}. As will be explained later, there is a an
obstacle deeply rooted in classical mechanics which has hindered CTMC
to be extended to more than one active target electron. One goal of
our work is to overcome this obstacle. In the present paper we have
restricted ourselves to three particles, namely electron impact
ionization of hydrogen, in order to see where the strengths and
weaknesses of our quasiclassical approach lie. Also, we would like to
make contact with the original CTMC method by deriving it as a
limiting case of our approach.

Combining the classical treatment with quantum effects has had a long
history in collision theory. Already shortly after Abrines and
Percival introduced the ``classical trajectory Monte-Carlo'' method
where the target electron was represented by an ensemble of Kepler
orbits of fixed energy (``microcanonical distribution''), alternative
descriptions were proposed, e.g., the Wigner transform of the quantum
wave function \cite{EIC81}, ``superpositions'' of microcanonical
distributions of different energies \cite{HAR83, COH85} or, only
recently, an ``optimum classical description'', based on symmetry
considerations and phase space partitioning \cite{RAK01}. In an
attempt to describe four-body systems helium was modeled classically
in an independent electron picture or the interaction between the two
bound electrons was switched off temporarily to prevent autoionization
(see, e.g. \cite{MCK87, OLS89, SCH92}). All these attempts had in
common, that they started from the purely classical model and tried to
extend it by implementing quantum elements (see, e.g. \cite{KIR80,
ZAJ86}). Such strategies have limited predictive power, as parameters
or switch-on times must be fitted to reproduce particular cross
sections.

As briefly reported in \cite{GEY01} we propose an alterative approach:
we start from the (exact) time-dependent quantum formulation,
translate it to phase space by means of the Wigner formalism and
finally approximate this description classically by setting $\hbar=0$.
Hence, we can approximate quantum collisions in a controlled way.
Furthermore, we gain additional freedom for the description of the
target electron. This leads to an improved agreement of quasiclassical
differential cross sections with the experiment and, even more
important, opens the way to incorporate multi-electron targets without
any further approximation to the dynamical description.

The paper is organized as follows: the derivation of the
``quasiclassical'' approximation will be given in section 2, in
section 3 we will characterize the four initial state distributions we
use for the hydrogen target; the results will be presented and
discussed in section 4. The last section contains a conclusion and an
outlook to multi-electron targets.


\section{The quasiclassical modeling of a collision}

We divide the complete ionization process into three parts:
setting up the initial state, propagating this initial phase space
distribution and finally extracting cross sections from the scattered
initial distribution. To arrive at a consistent formulation each
part is treated in the same way by translation into the Wigner
picture of quantum mechanics \cite{WIG32}. The classical approximation
is subsequently achieved by setting $\hbar = 0$ \cite{JOH87}.

\subsection{Initial distribution}

We construct the initial distribution $w(\vec{p},\vec{q})$ from the
$\hbar=0$ limit of the Wigner distribution. This distribution is
discretized and the individual mesh points in phase space then
represent initial conditions for solving classical equations of
motion. However, the resulting paths are not those of real electrons in real
space, rather they should be interpreted as the evolution of a
discretized phase space density which may have negative parts as well
(e.g., for the Wigner distribution). These parts simply contribute to
the cross sections with their negative weight, there is no need to
argue about ``negative probabilities'' which clearly demonstrates the
advantage of starting with a quantum formulation.

However, naively incorporating such a generalized initial phase space
distribution $w(\vec{p}, \vec{q})$ into the usual classical framework
leads to other difficulties \cite{EIC81}: in general
$w(\vec{p},\vec{q})$ will not be stationary under the classical
propagation, i.e., its Poisson bracket with the Hamiltonian of the
unperturbed target does not vanish, $\{H_0^i,w\}\ne 0$. Hence, the
initial target distribution will look very different at the time the
projectile has reached the target and the collision actually happens.

Another problem is even more severe: when using a classical phase
space distribution derived from a quantum wave function to select
initial values for a classical trajectory calculation, most of these
initial values do not have the energy of the hydrogen ground state,
but they start from a range of energies around the binding energy.
Consequently, when extracting cross sections, it is not sufficient any
more to only test the final energy of one of the electrons, as the
other electron's final energy can not be calculated as the difference to
the mean total energy. In a total cross section this energy spread
might be neglected, but cross sections differential in energy would
either be convoluted with the initial energy spread or, if the
final energy of both continuum electrons is added up for a specific
trajectory, only trajectories with those initial conditions from the
initial distribution contribute which initially started on the energy
shell. We will explicitly address this problem, which has not been
mentioned in previous calculations using non--microcanonical initial
distributions \cite{EIC81, HAR83, COH85}.

Next, we will propose a formulation of the propagation and the
extraction of the cross section that can deal with the difficulties
arising from general initial distributions.

\subsection{Propagation}

Quantum mechanical time dependent scattering is described by
calculating the transition amplitude between initial and final state
through the S--matrix, which is in turn related to the T--matrix
describing directly the cross section, see, e.g., \cite{TAY72},
\be{S-mat}
	S_{fi}= \langle f|\Omega^\dagger_{-}\Omega_{+}|i\rangle,
\ee \showlabel{S-mat}
where
\be{moel}
	\Omega_{\mp} = \lim_{t \to \pm \infty} U^{\dagger}(t)\,U_0(t)
\ee \showlabel{moel}
are the M\o{}ller operators. The meaning of $\Omega_{+}$, e.g., is to
propagate backwards with $U_{0}(t) = \exp[-i H_0^i t]$ using the
asymptotic initial Hamiltonian $H_{0}^{i}$ without the
projectile--target interaction and then forward under the full
Hamiltonian with $U(t)$. If the initial and final states are
eigenstates of the asymptotic Hamiltonians $H^i_0$ and $H^f_0$, often
a short version is used:
\be{S-simple}
	S_{fi}= \lim_{t\to\infty}\langle f|U(t)|i\rangle.
\ee \showlabel{S-simple}
By a Wigner transform the quantum time evolution operator
$U(t)$ can be directly transformed with the help of the quantum
Liouville operator $\mathcal{L}_q$, which reduces to the classical
Liouville operator $\mathcal{L}_c$ in the limit $\hbar\to 0$ \cite{HEL76}.
The latter describes the evolution of a phase space distribution
$w(\vec{p}, \vec{q},t)$ according to the Poisson bracket
\be{lio}
	\partial_{t}w = \{H,w\} \equiv -i\mathcal{L}_c w
\ee \showlabel{lio}
in  analogy to the quantum evolution of the density matrix $\rho$
generated by the commutator,
\be{lio2}
	\partial_{t}\rho= -i[H,\rho].
\ee \showlabel{lio2}
Hence, we could directly use the translation of \eq{S-simple} to
classical mechanics via the Liouville operator. In connection with the
microcanonical initial state distribution this is indeed equivalent to
the CTMC formulation \cite{KEL93, RAF85}.

However, \eq{S-simple} is insufficient, if the distribution is not
stationary under the initial asymptotic propagation. Instead, one must 
use the complete form \eq{S-mat} of the scattering operator $S =
\Omega^\dagger_{-}\Omega_{+}$. Correspondingly, the M\o{}ller operators
are translated into a classical propagation scheme, which
``transforms'' the initial non--stationary phase space distribution
$w_i(\gamma)$, where $\gamma = (\vec p_1,\vec q_1,\vec p_2,\vec q_2)$
is a point in the 12-dimensional phase space, into the ``scattered''
result of the reaction $w_f(\gamma)$ at the same point in time:
\be{moelclas}
	w_f(t\!=\!0)\, = \lim_{t \to +\infty}\lim_{t' \to -\infty}
	e^{i\mathcal{L}^{f}_ct}e^{-i\mathcal{L}_c(t-t')}
	e^{-i\mathcal{L}^{i}_ct'}w_i
	\,\equiv\, {\cal K}w_i(t\!=\!0)\,.
\ee \showlabel{moelclas}
To perform the actual calculation the initial distribution is
discretized, $w_i(\gamma) = \sum_n w_n\delta^{12}
(\gamma-\gamma_n^{i})$ with the normalization $\sum w_n = 1$. With
\eq{moelclas} we get as final distribution
\be{moelclas2}
	w_f(\gamma) = {\cal K}w_i = 
		\sum_nw_n\delta^{12}(\gamma-\gamma_n^{f}),
\ee \showlabel{moelclas2}
where each phase space point $\gamma_n^{f}$ emerges from
$\gamma_n^{i}$ through solving successively Hamilton's equations,
first with $H_0^{i}$, then with $H$, and eventually with $H_0^f$. With
this propagation scheme a non--stationary initial distribution will
spread when being propagated backwards with the asymptotic $\mathcal
L_{c}^{i}$. However, it will be refocused under the following forward
propagation with $\mathcal L_{c}$. Hence, when the actual collision
happens at $t\approx 0$ the original target distribution is restored,
slightly polarized by the approaching projectile.

The propagation now compensates for the non--stationarity of the
initial distribution, which means, that \emph{any} arbitrary phase
space description can be used for the target, even classically
unstable multi-electron targets like helium.

\subsection{Extraction of cross sections}
\label{sec:CrossSection}

According to \eq{S-mat} the cross section is extracted from the
overlap between the scattered initial wave function $S|i\rangle$ and
the asymptotic final state $|f\rangle$, which is an eigenstate of
$H_0^f$. For ionization we assume that $|f\rangle$ can be approximated
by two free electrons. However, before we come to the actual
evaluation we have to formulate the cross section such that it can
make full use of the non--stationary initial phase space distribution
$w_i(\vec p_1,\vec q_1)$, where ``1'' refers to the target electron.
Without modification the total energy $E$ of the final state forces by
energy conservation for each classical trajectory only those parts of
the initial phase space distribution to contribute to the cross
section which have the same total energy $E$ (see above). However,
this would bring us essentially back to the microcanonical
description. In order to make the entire non--stationary initial state
distribution ``visible'' to the collision process, we use the energy
transfer $\Delta E_1 = E^f_1 - E^i_1$ to the target electron rather
than its energy $E^f_1$ itself as a differential measure. Of course,
as long as the initial state is on the energy shell with a well
defined energy $E = E^i_1 + E^i_2$ the new definition coincides with
the usual expression for the cross section,
\be{CrossSec}
	\left. \frac{d^5\sigma}{d\Omega_1\, d\Omega_2\, dE_1}\right |_{E} =
	\left. \frac{d^5\sigma}{d\Omega_1\, d\Omega_2\, d\Delta 
E_1}\right |_{E}\,,
\ee \showlabel{CrossSec}
where $d\Omega_i$ are the  differentials for the solid angles of the
two electrons, respectively.

To extract this cross section  we have to  evaluate the phase space
integral \cite{ROS98, CAR83}
\be{PhaseSpaceInt}
\noindent
\frac{d^5\sigma}{d\Omega_1\,d\Omega_2\,d\Delta E_1}=
	\int \!\! d\gamma^i\,
	\delta(\Omega_1^f\!-\Omega_1)\, \delta(\Omega_2^f\!-\Omega_2)\,
	\delta(\Delta E_1^f\!-\Delta E_1)\: w_f,
\ee \showlabel{PhaseSpaceInt}
where the integration is over the initial state variables with the
``scattered'' distribution $w_f(\gamma^f) = \mathcal{K}w_i(\gamma^i)$.

The cross section (\ref{PhaseSpaceInt}) is a generalization of the one
derived, e.g., in \cite{ROS98}, where the initial target bound state
was assumed to live on a torus, i.e., $w_i(\vec p_1,\vec q_1) = \delta
(\vec I(\vec p_1,\vec q_1)-\vec I_0)$ with a well defined
multidimensional action $\vec I_0$ and the initial project state fixes
$(\vec{p}_2, \vec{q}_2)$, except for the impact parameter area $dx_2
dy_2$. Note that the change of variables from the momenta and
positions for the scattered distribution $w_f$ to scattering angles
and energy transfer in \eq{PhaseSpaceInt} contains a constraint how a
differential phase space volume centered around its guiding trajectory
contributes to the cross section: only those phase space cells are
relevant for ionization whose trajectory has the correct energy
transfer \emph{and} positive energy after the change of variables.
Trajectories with correct energy transfer but final negative energy
(i.e., one electron is still bound) do not contribute to the
ionization cross section. Consequently, if one is interested in the
total ionization cross section only, \eq{PhaseSpaceInt} simplifies to
\be{eq:sigma-tot}
	\sigma_{tot} = \pi b_{max}^2 \sum_{E_1^f, E_2^f \geq 0}
	w_n \:\Theta(\Delta E_1^f\! -\Delta E_1)\,.
\ee \showlabel{eq:sigma-tot}

Finally, we have to respect the Pauli principle for the two identical
electrons. This can be done in the initial or final state. The Wigner
transformed (anti)--symmetrized asymptotic state wave function has
two parts: the ``classical'' part, independent of $\hbar$, where the
indices of the two electrons are interchanged and an ``interference
term''. In the limit $\hbar\to 0$ the latter part oscillates
infinitely rapidly and its contribution to the cross section
\eq{PhaseSpaceInt} vanishes upon integration. In the actual
calculation it is easier to perform the remaining classical
symmetrization in the final state.

\section{Initial target phase space distributions}

In the previous section we have taken the theoretical effort to
formulate an approximate scattering theory which goes beyond the
classical approximation still using classical trajectories as in CTMC.
Our goal was to allow for non--stationary initial target distributions
in phase space as they occur if one translates the quantum wave
function into a phase space distribution. This translation, however,
is not unique and depends on the used correspondence rule as shown by
Moyal \cite{MOY49}, see also \cite{MEH64, SHE59, COH66}.

Our main interest in this first application of non--stationary
distributions is not to optimize the initial distribution but to learn
how the collision dynamics is influenced by different aspects of
initial distributions. For this purpose we have chosen four prominent
distributions for which we will compare the same collision processes.
Apart from the well known microcanonical distribution we have chosen
Cohen's ensemble, which is a superposition of microcanonical
distributions \cite{COH85}. In addition to these more classical
distributions we will implement the product distribution, the product
of the quantum mechanical density in momentum and coordinate space
\cite{COH66}, and the well known Wigner distribution \cite{WIG32}.

All four distributions have the correct expectation value of the
binding energy of $E_b = -0.5$ a.u.> However, they have different energy
spreading, defined as
\be{eq:DiscrESpread}
	(\Delta E)^2 = \frac{
	\int d^3p\,d^3q\; w(\vec{p},\vec{q})\: (H(\vec{q}, \vec{p}) - E_b)^2}
		{\int d^3p\,d^3q\; w(\vec{p},\vec{q})}.
\ee \showlabel{eq:DiscrESpread}

>From table \ref{tbl:ESpreads} one sees that the Wigner and the product
distribution have a rather broad energy distribution. This is in
general the price one must pay for phase space distributions which
resemble closely quantum wave functions. For later reference we
characterize very briefly each distribution.

\begin{table}[h]
	\centering
	\begin{tabular}{c|cccc}
		distribution & microcan. & Cohen & product & Wigner \\
		\hline \\[-0.3cm]
		$\sqrt{(\Delta E)^2}$ [a.u.] & 0 & 0.24 & 1.22 & 0.88 \\
	\end{tabular}
	\caption{Energy spread of the phase space distributions used in
	our calculations, see text.}
	\label{tbl:ESpreads}
\end{table}
\showlabel{tbl:ESpreads}

\subsection{The Wigner distribution}
\label{sec:WignerDist} \showlabel{sec:WignerDist}

 From a wave function $\psi(\vec r)$ one obtains the Wigner distribution
by the reversible transformation \cite{WIG32}
\be{eq:Wignertrafo}
	w(\vec p, \vec q; t) =
	\frac{1}{(2\pi\hbar)^3}\int\!\! d^3s \,
	e^{i\vec{p}\vec{s}/\hbar} \psi^*(\vec{q}+\vec{s}/2, t)\:
	\psi(\vec{q}-\vec{s}/2, t).
\ee \showlabel{eq:Wignertrafo}
By construction the Wigner function reproduces the quantum probability
densities in coordinate and momentum space:
\be{eq:q-p-densities}
	\int \!\! d^3p \, w(\vec p, \vec q) = | \psi(\vec q)|^2
	\quad \mbox{and} \quad
	\int \!\! d^3q \, w(\vec p, \vec q) = | \psi(\vec p)|^2\,.
\ee \showlabel{eq:q-p-densities}
Note that the Wigner function depends on the angle between $\vec{p}$
and $\vec{q}$ with negative contributions outside the classically
allowed region, thus incorporating quantum correlations into phase
space. We have calculated the Wigner function following Eichenauer
etal. \cite{EIC81}.

\subsection{The product distribution}

As the name indicates the product distribution is defined as the
product of the radial quantum probabilities in coordinate and momentum
space \cite{COH66}, in our case for the hydrogen ground state:
\bea
	w(r,p) = \rho(r) \times \sigma(p) \quad \mbox{with}
		\nonumber\\
	\rho(r) = 4r^2e^{-2r} \quad \mbox{and} \quad
		\sigma(p) = \frac{32p^2}{\pi(1+p^2)^4},
		\label{eq:QuantumRDens}
\eea
\showlabel{eq:QuantumRDens}
where $r=|\vec{q}|$ and $p=|\vec{p}|$. Clearly, by definition the
correct quantum density is obtained in coordinate or momentum space as
for the Wigner function \eq{eq:q-p-densities}. However, now the
directions of $\vec{q}$ and $\vec{p}$ are independent of each other.

\subsection{The microcanonical distribution}

The microcanonical distribution is the standard (classical)
distribution of CTMC based on Bohr's model of the hydrogen atom
\cite{ABR66a}. The classically accessible area of phase space on the
energy shell is sampled with equal probability in angle--action
coordinates. This distribution has the ``correct'' vanishing spread of
the binding energy and the correct quantum density in momentum space.
However, the coordinate space density
\be{eq:MicroRDens}
	\rho_{\mbox{\scriptsize micro}}(r) = 
\frac{2r^2}{\pi}\sqrt{\frac{2}{r}-1},
\ee \showlabel{eq:MicroRDens}
has a cutoff which is not present in the quantum density (cf.
\eq{eq:QuantumRDens}). Since the microcanonical distribution is
derived from Bohr's atomic model, there exists, of course, no
extension to construct a corresponding classical distribution for more
complex targets, e.g., the helium atom. Finally we note that the
Wigner distribution, if forced to be on the energy shell, reduces to
the microcanonical distribution for hydrogen.

\subsection{Cohen's distribution}

In order to preserve the stationarity of the microcanonical
distribution, but still come closer to the quantum densities Cohen
proposed another distribution which we will call Cohen's distribution
in the following \cite{COH85}: It is obtained by summing
microcanonical distributions of different energies such that the mean
binding energy is correct and only bound orbits contribute. The
resulting (analytical) coordinate density is set to the quantum
mechanical one. By these constraints the system of equations is
already fully determined and the momentum density obtained is only
slightly different from the quantum mechanical one. The energy spread
of this distribution is small compared to the Wigner or product
distribution.

In the following we will see how these four distributions perform
under different collisions. Thereby, we will also gain some insight
which properties of the target system are highlighted by a specific
cross section.


\section{Results for electron impact ionization of hydrogen
at three different impact energies}

We will present our results for the three different impact energies of
250 eV, 54.4 eV and 17.6 eV. These values are motivated by existing
experimental data for comparison. At a given impact energy all cross
sections, total, single and multiple differential ones, can be
extracted from the same numerical data set, see table
\ref{tab:NumOfTrajs}. The only difference is that higher differential
cross sections require many more trajectories to achieve sufficient
statistics since only a few hundred out of several million
trajectories typically contribute to a fully differential cross
section.

\begin{table}[h]
	\centering
	\begin{tabular}{c|c|ccc}
		$E_{in} [eV]$ & distribution & $b_{max}$ [a.u.] & $N_{tot}$
			& $N_{ion}$  \\
		\hline \\[-0.3cm]
		250 & microcan. & 3.5 & 6.1 Mio. & 155'484  \\
		    & product & 12 & 186 Mio. & 395'691  \\
		    & Wigner & 15 & 104 Mio. & 339'242  \\
		    & Cohen & 7.5 & 17.75 Mio. & 83'492 \\
		\hline
		54.4 & microcan. & 6 & 20 Mio. & 455'781  \\
		     & product & 22 & 80 Mio. & 162'752  \\
		     & Wigner & 22 & 181 Mio. & 537'997  \\
		     & Cohen & 9 & 25.67 Mio & 222'888 \\
		\hline
		17.6 & microcan. & 4.5 & 17.5 Mio. & 320'244 \\
		     & product   & 9   & 16.5 Mio. & 79'146  \\
		     & Wigner    & 22  & 50.5 Mio. & 47'077  \\
		     & Cohen     & 8   & 14.9 Mio. & 80'780 \\
	\end{tabular}
	\caption{Overview of the numbers of trajectories calculated for
	the various impact energies and initial distributions: $b_{max}$
	denotes the maximum impact parameter used, $N_{tot}$ and $N_{ion}$
	give the total numbers of trajectories and those contributing to
	ionization, respectively (cf. section \ref{sec:CrossSection}).}
	\label{tab:NumOfTrajs}
\end{table}
\showlabel{tab:NumOfTrajs}

For this reason we have carefully implemented the numerical
integration of the trajectories. The equations of motion have been
regularized to avoid the attractive Coulomb singularities \cite{KUS65}
and a sixth order symplectic integrator \cite{YOS90} with adaptive
stepsize control has been used. Hence, we are able to handle electron
trajectories which directly hit the nucleus. As a consequence the
integration is very stable: At a given total energy less than ten
trajectories had to be discarded due to a too large integration error.
This stability is important, since the trajectories discarded should
be significantly fewer than those contributing to a fully differential
cross section.

\subsection{Total ionization cross sections}

For the microcanonical distribution, the product distribution and
Cohen's energy distribution every trajectory $j$ has the same weight
$w_j = +1/N_{tot}$ in \eq{eq:sigma-tot} which reduces as a consequence
to
\be{eq:totion}
	\sigma_{tot} = \pi b_{max}^2 N_{ion}/N_{tot}\,,
\ee
the familiar form of standard CTMC. The result for the three impact
energies under consideration is given in table \ref{tab:sigma-tot}.

\begin{table}[h]
	\centering
	\begin{tabular}{c|ccc}
		distribution & $E_{in}$ = 250 eV & 54.4 eV & 17.6 eV  \\
		\hline \\[-0.3cm]
		microcan.  & 2.76 & 7.24 & 3.27 \\
		product    & 2.71 & 8.69 & 3.55 \\
		Wigner     & 2.52 & 7.91 & 3.98 \\
		Cohen      & 2.33 & 6.20 & 3.06 \\
		experiment & 3.43 & 6.19 & 2.1\\
	\end{tabular}
	\caption{Total ionization cross section in $10^{-17}cm^2$ at the
	three main impact energies compared to experimental data by Shah
	etal. \cite{SHA87}.}
	\label{tab:sigma-tot}
\end{table}
\showlabel{tab:sigma-tot}

\epsfxsize=13cm
\epsfysize=4.5cm
\begin{figure}[htb]
	\hspace{0.5cm}
	\epsffile{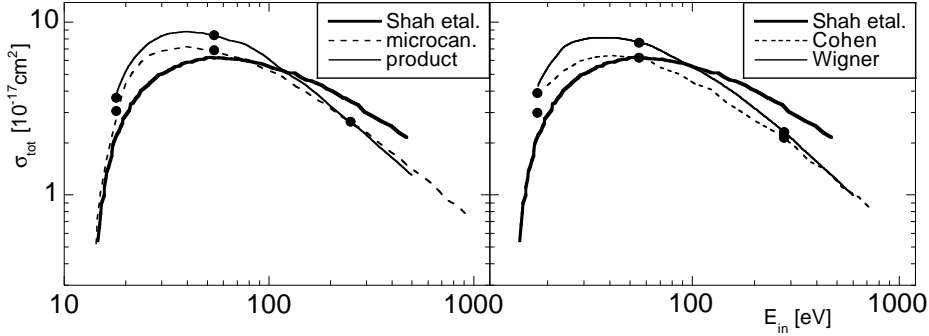}
	\caption{Absolute total ionization cross section calculated with
	the four compared initial distributions. The measurements (thick
	solid line) are from by Shah etal. \cite{SHA87}. The dots denote
	the energies, where the differential cross sections were
	calculated.}
	\label{fig:sigma-tot}
\end{figure}
\showlabel{fig:sigma-tot}

The total cross section was calculated for the energy range from 15 eV
up to 500 eV impact energy. As one can see from \fig{fig:sigma-tot}
the overall trend of the experimental cross section is reproduced.
However, the energy at the maximum of the cross section is too low by
nearly a factor of two and the high energy limit of the calculations
obeys the classical $1/E$ law \cite{THO12} instead of the quantum
mechanical $\ln(E)/E$ behavior \cite{BET30}. The fact that all initial
state distributions, from the classical microcanonical one to the
quantum Wigner distribution, result in the same total cross section to
within some ten percent shows that the deviation from the exact result
is mainly due to the dynamical evolution of the distribution during
the collision. The classical evolution cannot account for quantum
tunneling, e.g., which becomes relevant for higher impact energies.

Can one hope under these circumstances to obtain reasonable results
for differential cross sections? As it has been shown previously
\cite{BRA89} disagreement in the absolute values of the cross section
does not necessarily imply a similar disagreement in the shape of the
differential cross sections.

\subsection{Single differential ionization cross section}

\begin{figure}[htb]
	\centering
	\hspace{0.5cm}
	\epsffile{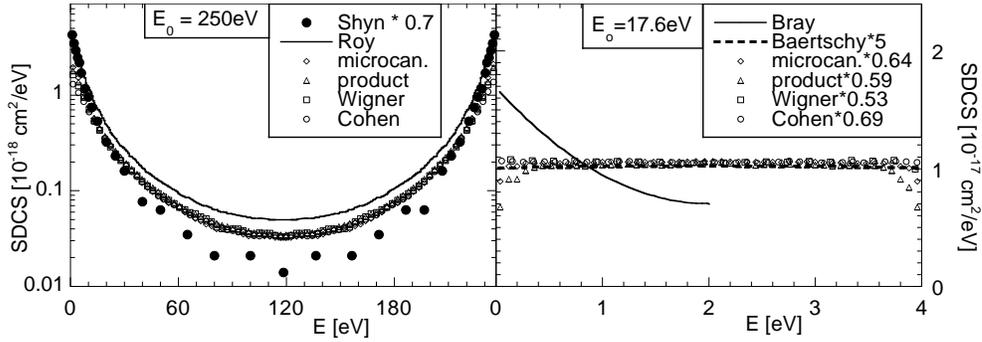}
	\caption{Singly differential cross section $\frac{d\sigma}{dE}$
	for different initial distributions at $E_{in}$ = 250 eV (left)
	and at $E_{in}$ = 17.6 eV (right). The measurements by Shyn
	\cite{SHY92} at 250 eV and our calculations at 17.6 eV are scaled
	to the correct total cross section. The quantum calculation (solid
	line) is from Roy \cite{ROY01}. Comparison at 17.6 eV is made with
	calculations by Bray \cite{BRA00a} and Baertschy etal.
	\cite{BAE01a}, see text.}
	\label{fig:SDCS250-17}
\end{figure}

Generally the single-differential cross section evolves from a
U--shaped form for high energies (\fig{fig:SDCS250-17}, left panel, at
250 eV) to a flat curve for low energies (\fig{fig:SDCS250-17}, right
panel, at 17.6 eV) The cross section at 54.4 eV follows this trend and
is not shown here. There is little difference between the initial
distributions. The agreement with the (scaled) experiment by Shyn
\cite{SHY92} as well as with a calculation by Roy \cite{ROY01}
employing an exchange--modified Glauber approximation is fairly good
at 250 eV. Our result is also consistent with a coupled pseudo state
approximation \cite{CUR87} and a distorted-wave Born calculation
\cite{MCC89}. At 17.6 eV the form of the quasiclassical cross section
agrees well with results for the $L=0$ partial wave obtained with two
different fully quantal methods, namely exterior complex scaling and a
time dependent CCC method \cite{BAE01a}. However, there is clear
disagreement with Bray's CCC calculation \cite{BRA00a}. In this
context it should be noted that towards threshold (13.6 eV impact
energy) the contributions of all partial waves should behave as the
S-wave contribution \cite{ROS95}. Moreover, the Wannier theory and
classical calculations \cite{ROS94} predict a flat cross section to
within 5\%.

Overall, the single differential cross sections are in good agreement
with the experiment and full quantum calculations and they do not
provide a critical test for the used initial phase space
distributions. However, the very formulation of energy differential
cross sections for non--stationary initial distributions has been
possible only by the definition of the cross section in terms of the
energy transfer (see sec. \ref{sec:CrossSection} and table
\ref{tbl:ESpreads}).

\epsfxsize=13cm
\epsfysize=4.5cm
\showlabel{fig:SDCS250-17}

\subsection{Multiply differential cross sections at 250 eV impact energy}

The impact energy of 250 eV has often been studied, since it is high
enough for the first Born approximation to be applicable, but still
not too high for a reasonable counting rate in the experiments.

\subsubsection{Double differential ionization cross section}
\label{sec:DDCS250} \showlabel{sec:DDCS250}

\begin{figure}[htb]
	\centering
	\hspace{0.5cm}
	\epsffile{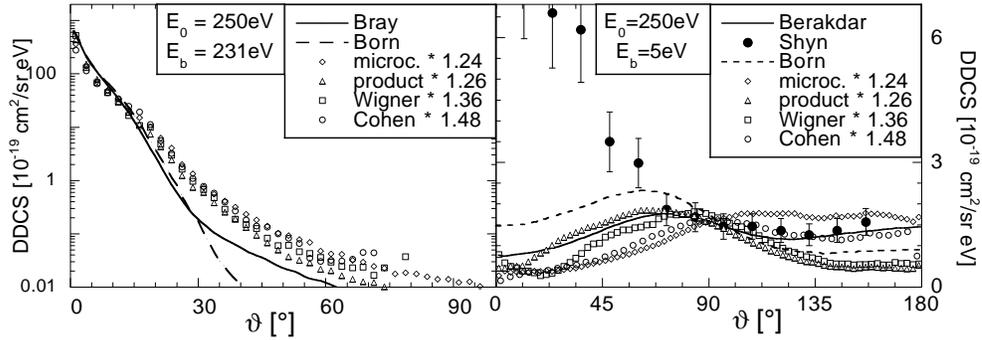}
	\caption{Doubly differential cross section for the fast ($E_b$ =
	231.4 eV, left panel) and the slow electrons (5 eV, right panel).
	The DDCS for the different initial distributions are scaled to
	give the correct total cross section. The cross sections for the
	fast electrons are compared to the quantum calculations of Bray
	\cite{BRA00b}, the slow electrons to the calculation of Berakdar
	\cite{BER93} and the experiments of Shyn \cite{SHY92}. ``Born''
	denotes a first order Born approximation.}
	\label{fig:DDCS250}
\end{figure}

As an example for a double differential cross section we present the
angular distribution of the electrons for fixed energy sharing of
231.4 eV + 5 eV. The shape from the different distributions in the
left panel of \fig{fig:DDCS250} for the fast electron is similar and
close to the quantum calculations (there are no experiments available
for this energy).

\epsfxsize=13cm
\epsfysize=4.5cm
\showlabel{fig:DDCS250}

The situation is quite different for the slow electron
(\fig{fig:DDCS250}, right panel). However, the large deviation from
the experiment at small angles is probably due to a systematic error in
the experiment \cite{SHY92} (see remarks in \cite{BER93, BRA00b}).
More interesting for us is the fact that already at the level of double
differential cross sections different initial state distributions
begin to make a difference. In particular the microcanonical
distribution leads to a cross section with opposite trend compared to
the others. The binary peak (around 70$^\circ$) is almost absent and
the cross section is shifted towards larger angles. Apparently, the
angular dependence of the scattered slow electron is sensitive to the
initial spatial distribution of the bound electron which is not
correctly modeled in the microcanonical distribution (see
\eq{eq:QuantumRDens}, \eq{eq:MicroRDens}).

\subsubsection{Triple differential ionization cross section}

For electron impact ionization of hydrogen at an impact energy of 250
eV a wide range of available data exists from measured fully
differential cross sections (see, e.g. \cite{EHR85, EHR86, WEI79,
LOH84}) as well as from calculated ones,  for a recent
review see \cite{LUC99}.

The small volume of the final phase space imposes a serious
statistical limit on our quasiclassical calculations: A triply
differential cross section at the experimental resolutions of about
$\pm$ 1$^\circ$ in the scattering angle of the fast electron, $\pm$
2$^\circ$ for the slow electron, the definition of the scattering
plane at $\pm$ 10$^\circ$ and an energy resolution of less then 1 eV
consists of a fraction of less then $10^{-9}$ of the final state phase
space volume only. If the ionization events were equally distributed
over the whole phase space, we would hardly find one trajectory for a
cross section. Due to the structure in the pattern of the outgoing
electrons, we can record about 100 to 300 events for those
configurations where experimental data is available.

The small number of trajectories contributing to the TDCS forced us to
set the bins of the scattering angle $\theta_b$ larger than their
distance. With this trick each event contributes to two or three
adjacent data points of the cross section. Still, the comparison with
the DDCS shows that the number of propagated trajectories would have
to be increased by one or two orders of magnitude for full statistical
convergence.

Once the trajectories have been calculated it is easy to filter out
cross sections for arbitrary final electron configurations. We will
show cross sections for some of the measured data in Ehrhardt
geometry, where the fast electron is fixed at a small angle and the
angular distribution of the slow electron at a fixed energy is
measured: This configuration is very sensitive to the details of the
initial distribution and, due to the slow ionized electron, to the
dynamics of the reaction. In general two main features appear: the so
called binary peak, located around the direction of the momentum
transfer (in the range between 70$^{\circ}$ and 85$^{\circ}$) is
usually explained by the classical picture of a binary scattering of
the two electrons without interacting with the nucleus. The recoil
peak consists of electrons which are rescattered from the nucleus by
180$^{\circ}$ after the binary encounter. The recoil peak lies roughly
opposite to the binary peak (for an overview over the various
processes and their explanations see, e.g., \cite{BRI89}).

\begin{figure}[htb]
	\centering
	\hspace{0.5cm} \epsffile{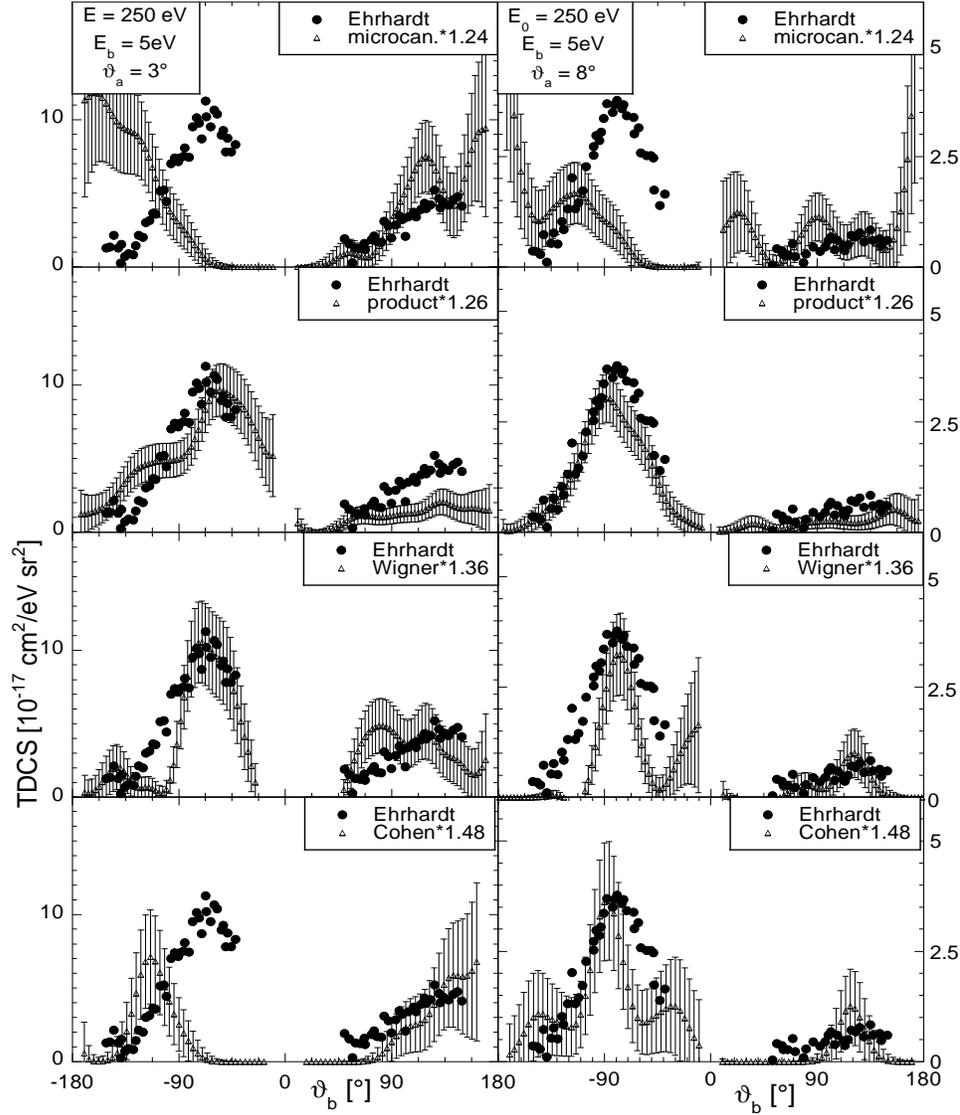}
	\caption{Fully differential cross sections in Ehrhardt geometry.
	The present calculations are scaled to give the correct total
	cross section. The absolute measurements are from Ehrhardt etal.
	\cite{EHR85}.}
	\label{fig:TDCS250-5}
\end{figure}

Out of the many parameters for which experimental data is available we
show two examples from Erhardt \cite{EHR85, EHR86} in figure
\ref{fig:TDCS250-5}. Angles out of the scattering plane between $\pm
5^\circ$ and $\pm 20^\circ$ have virtually no influence on the shape
of the cross section, they only affect the statistics. We set the
angular range to $\pm 10^\circ$.

\epsfxsize=13cm
\epsfysize=14.8cm
\showlabel{fig:TDCS250-5}

As it is clear from figure \ref{fig:TDCS250-5} the microcanonical
distribution is not capable of reproducing the main features of the
cross section: there is no binary peak (the ``classical'' part of the
cross section) at small momentum transfer and at higher momentum
transfer it is much too small and shifted to larger angles. There is
some structure in the region of the recoil peak, but as its maximum is
opposite to the direction of the fast outgoing electron, this is
probably an ``artifact'' of the calculation.

To describe the recoil peak in a quantum Born approximation the final
state wave function for the slow electron has to be a Coulomb wave,
i.e., it must include the interaction with the nucleus. The overlap of
this infinitely extended wave function with the small, localized,
radially symmetric ground state selects only the central region of the
Coulomb wave, which correspondents to rescattering at vanishing impact
parameter. In the classical approximation only a vanishingly small
fraction of trajectories leads to a head on collision with the nucleus
after the momentum has been transferred. Hence the probability for
rescattering by 180$^\circ$ is negligible and this quantum
interpretation of the recoil peak can only be used with caution for
the classical approximation. However, one should keep in mind that
this picture of the electron recoil from the nucleus comes from the
Born perturbation series with its hierarchy of orders, i.e., a number
of sudden two particle interactions takes place one after the other
\cite{BRI89}. It is only in this interpretation, that ionization and
rescattering are two distinct events; in the full dynamical
description all three bodies interact simultaneously.

The other three distributions perform remarkably better, especially
with the product and the Wigner distribution the position and the
width of the binary maximum are reproduced to within the statistical
uncertainties. These two distributions are characterized by the
correct densities in coordinate \emph{and} momentum space, whereas
Cohen's energy distribution lacks the correct momentum distribution
and the microcanonical one has the wrong spatial dependence. We may
conclude, that at 250 eV impact energy the cross section is a
projection of the initial distribution's phase space density onto the
final configuration: both, in momentum and in coordinate space the
description of the target must be correct.

\subsection{Multiply differential cross sections at 54.4 eV impact energy}

The impact energy of 54.4 eV is close to the maximum of the
experimental total cross section. The presentation of the cross
sections will be similar to that at 250 eV.

\subsubsection{Double differential ionization cross section}

\begin{figure}[htb]
	\centering
	\hspace{0.5cm}
	\epsffile{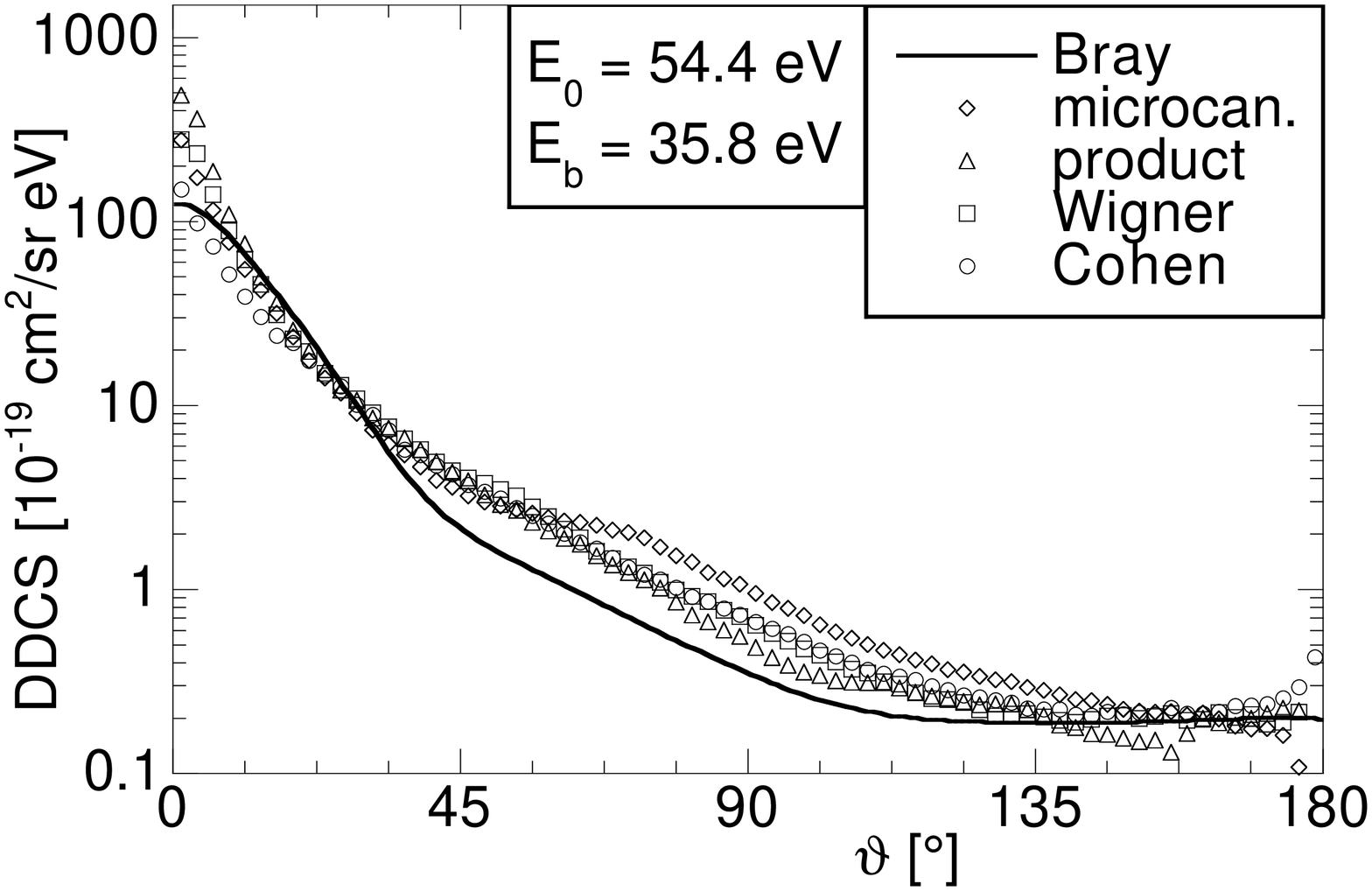}
	\caption{Doubly differential cross section at $E_{in}$ = 54.4 eV
	of the fast electrons with $E_b$ = 35.8 eV: comparison of the
	quasiclassical results with a CCC calculation by Bray
	\cite{BRA00b}.}
	\label{fig:DDCS54-36}
\end{figure}

\begin{figure}[htb]
	\centering
	\hspace{0.5cm}
	\epsffile{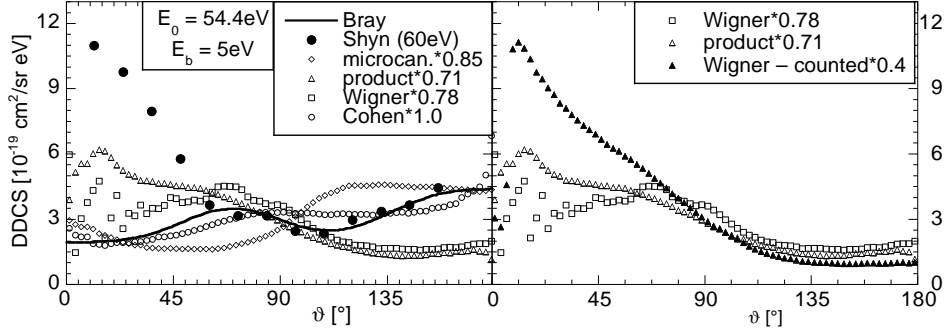}
	\caption{Left: Doubly differential cross section at $E_{in}$ =
	54.4 eV of the slow electrons with $E_b$ = 5 eV. The
	quasiclassical and the experimental results were scaled to the
	correct total cross section. The experiment \cite{SHY92} is wrong
	again at small angles, the full curve is the fully quantum
	mechanical calculation of ref. \cite{BRA00b}. Right: the same as
	left, but here the Wigner result is \emph{not} weighted with the
	sign of the Wigner distribution, see text for explanation.}
	\label{fig:DDCS54-5}
\end{figure}

The doubly differential cross section is presented again for a pair of
matching energies for the fast (\fig{fig:DDCS54-36}) and the slow
electrons (\fig{fig:DDCS54-5}) with 35.8 eV and 5 eV, respectively. No
experimental data exists for the fast electrons. Comparison, however,
is possible with Bray's CCC calculation \cite{BRA00b}. For the slow
electrons, experiments by Shyn are available, but only at an impact
energy of 60 eV. These experimental data were scaled onto the correct
total cross section at 54.4 eV. They show systematically too high
values at small angles similarly as for 250 eV (cf. section
\ref{sec:DDCS250} and refs. \cite{BER93, BRA00b}).

\epsfxsize=6.5cm
\epsfysize=4.5cm
\showlabel{fig:DDCS54-36}

The cross sections for the faster electrons are similar
(\fig{fig:DDCS54-36}), the quasiclassical results show slightly higher
contributions at angles above 45$^{\circ}$, almost independent of the
phase space distribution describing the target.

\epsfxsize=13cm
\epsfysize=4.5cm
\showlabel{fig:DDCS54-5}

For the slow electrons the picture is quite different
(\fig{fig:DDCS54-5}, left panel): the Wigner distribution reproduces
the position of the binary peak (about 70$^\circ$) and its width
fairly well, whereas the microcanonical distribution nearly inverts
the shape of the quantum mechanical cross section. Cohen's energy
distribution gives a cross section somewhere between the quantum
Wigner description and the classical microcanonical modeling.

The cross section from the product distribution is much too high at
small angles. This behavior can be simulated with the Wigner
distribution as well, if $|w_{n}|$ is taken instead of $w_{n}$ in
\eq{eq:sigma-tot}, see the right panel of \fig{fig:DDCS54-5}.

The Wigner and the product distribution both reproduce the correct
density in coordinate and momentum space; the difference is that in
the quantum Wigner distribution the radius and the momentum vector are
correlated with negative weights in regions outside of the classical
turning point (see sec. \ref{sec:WignerDist}). The difference between
the Wigner cross section with the correct sign for each final value
and the one with the same sign for each trajectory shows, that the
contributions at small scattering angles come from these outer
regions. The negative parts of the Wigner distribution lead to partial
cancellation of classically allowed contributions, which nevertheless
do not occur in the cross section. The product distribution lacks the
correlations and consequently gives the wrong cross section at small
angles, here all trajectories contribute. In the microcanonical
distribution no initial values at all exist for radii larger than 2
a.u., consequently the cross section is much too small for small
angles. We conclude that this type of a doubly differential cross
section strongly depends on quantal features which are beyond the
scope of a purely classical ansatz as standard CTMC. However, they can
be incorporated in our $\hbar=0$ approximation through the initial
distribution.

\subsubsection{Triple differential ionization cross section}

The triply differential cross section at 54.4 eV impact energy has been
measured by R\"oder etal \cite{ROD96} in Ehrhardt geometry. We show
the cross sections for the parameters given in \tbl{tbl:TDCSParams54}.

\begin{table}[h]
	\centering
	\begin{tabular}{c|cc}
		$E_b$ [eV] & $\theta_a$ & measured by  \\
		\hline \\[-0.3cm]
		5 $\pm$ 0.5 & 4$^\circ$ $\pm$ 1$^\circ$, 23$^\circ$ 
$\pm$ 2$^\circ$
		& R\"oder etal. \cite{ROD96} \\
	\end{tabular}
	\caption{Overview over the parameters for the triply differential
	cross sections in Ehrhardt geometry extracted at $E_{in}$ = 54.4
	eV. The angle between the scattering planes was confined to $\pm
	12^{\circ}$.}
	\label{tbl:TDCSParams54}
\end{table}
\showlabel{tbl:TDCSParams54}

The DDCS for the fast electrons, \fig{fig:DDCS54-36}, is too high for
small angles. We therefore not only scale the TDCS onto the total
cross section, but additionally for $\theta_a$ = 4$^\circ$ by a factor
of 0.25. This scaling has been performed to see, if the shape of the cross sections
for the slow electrons is reproduced correctly.
\epsfxsize=13cm
\epsfysize=14.8cm
\begin{figure}[htb]
	\centering
	\hspace{0.5cm} \epsffile{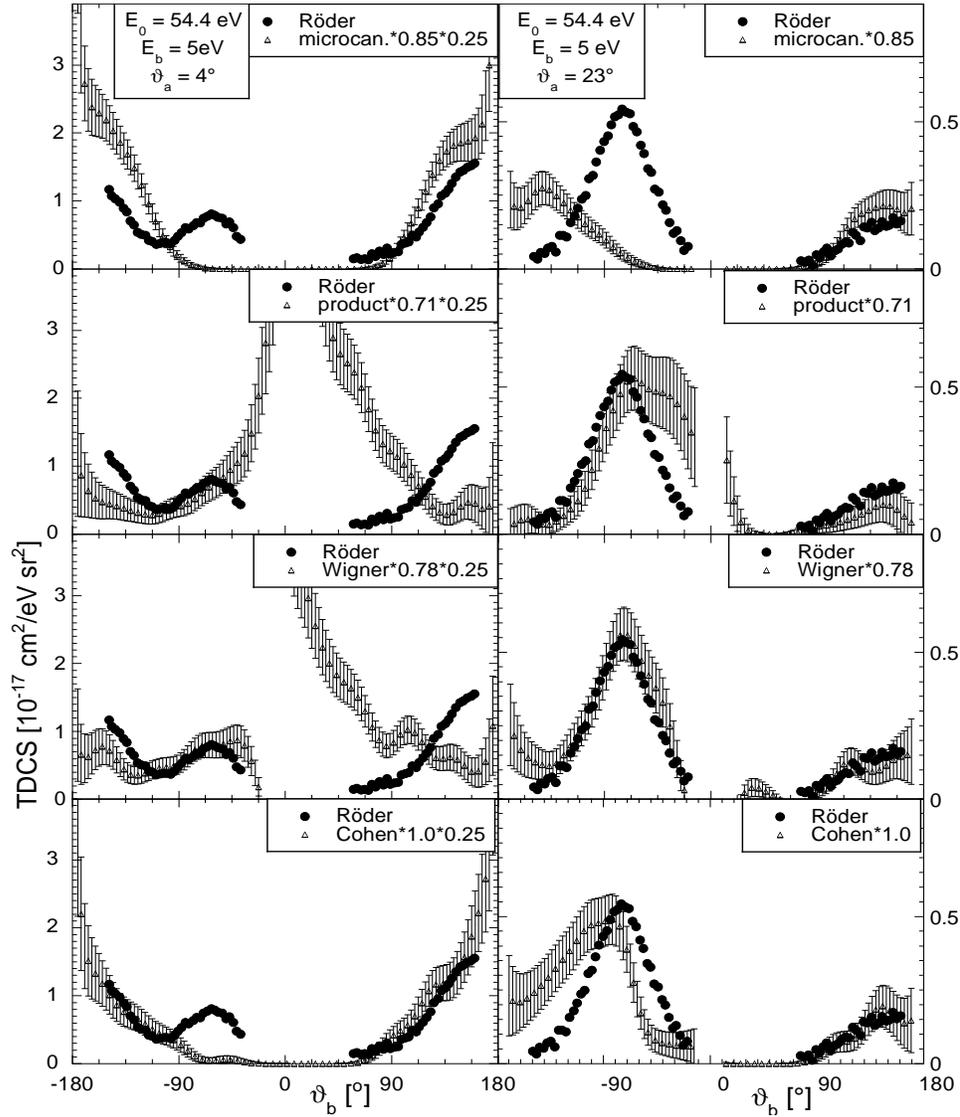}
	\caption{Fully differential cross sections in Ehrhardt geometry at
	54.4 eV impact energy for a fixed angle of $\theta_a$ = 4$^\circ$
	(left) and 23$^\circ$ (right). The calculations are scaled
	according to the total cross section and the form of the DDCS (for
	4$^\circ$, see text). The absolute measurements are by R\"oder
	etal. \cite{ROD96}}
	\label{fig:TDCS54-423}
\end{figure}
\showlabel{fig:TDCS54-423}

The differences between the phase space distributions can be seen
clearly: the classical microcanonical distribution fails to reproduce
the binary peak at all parameters, whereas the recoil peak agrees
astonishingly well. The next more quantal description, Cohen's energy
distribution, performs better with increasing $\theta_a$, but the
angle of the binary peak is always too large. The opposite tendency,
placing the binary peak at too small angles, is seen with the product
distribution. At small $\theta_a$ the electrons even leave in the same
direction, a configuration that should be strongly suppressed by their
mutual repulsion. This problem will be discussed in detail for the
impact energy of 17.6 eV. The distribution which is closest to quantum
mechanics, the Wigner distribution, can reproduce the cross sections
over the entire range of parameters, but with similar effects of both
electrons leaving in forward direction at $\theta_a$ = 4$^\circ$.

At this impact energy switching from the classical models of the
microcanonical or Cohen's energy distribution that only reproduce
either the coordinate or the momentum space density, to the more
quantal ones results in a definite improvement in the cross sections.
The product distribution, which satisfies both momentum and coordinate
space densities, already shows a binary peak, although shifted. With
the correlated Wigner distribution the best correspondence between
experiment and our calculations is achieved.

\subsection{Multiply differential cross sections at 17.6 eV impact energy}

At the low impact energy of 17.6 eV the projectile is only slightly
faster than a classical electron on a circular orbit with the Bohr
radius and the ionized electrons can only share 4 eV of energy. Hence,
we expect different qualities of the initial distribution to become
important, compared to the higher impact energies.

At these low energies quantum calculations, using analytic
final state wave functions constructed from two-body wave functions,
fail or perform only very poorly \cite{LUC99}. This indicates, that
here a correct description of the simultaneous interaction of the three
particles is crucial. In this context it should be emphasized that we do not
approximate the three-body dynamics beyond the classical
$\hbar=0$ limit: All long range Coulomb interactions
between all particles  are taken fully into account.

\subsubsection{Double differential ionization cross section}

For the DDCS no experimental comparison exists, our results will be
compared to Bray's CCC calculation \cite{BRA00a} and to results of an
ECS calculation by Isaacs etal \cite{ISA01}. For symmetric energy
sharing the CCC single differential cross section is too small by
approximately one third, hence the CCC DDCS was scaled by $1/0.66$ for
better comparison of the shape. Our results are scaled onto the total
cross section again.

\epsfxsize=13cm
\epsfysize=4.5cm
\begin{figure}[htb]
	\centering
	\hspace{0.5cm}
	\epsffile{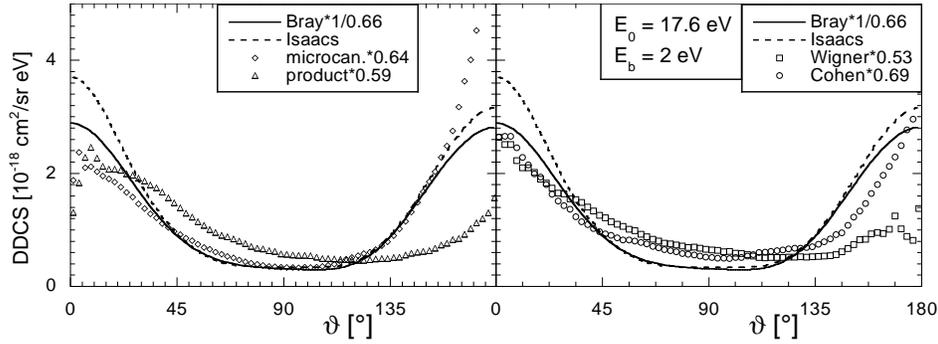}
	\caption{Angular distribution of the electrons with $E_b$ = 2 eV
	at $E_{in}$ = 17.6 eV, scaled onto the total cross section,
	compared to quantum calculations by Bray \cite{BRA00a} and Isaacs
	\cite{ISA01} etal. For the scaling see text and
	\fig{fig:SDCS250-17}.}
	\label{fig:DDCS17}
\end{figure}
\showlabel{fig:DDCS17}

As can be seen in \fig{fig:DDCS17}, the more classical phase space
distributions give considerably better results. With the
microcanonical distribution the cross section deviates only for angles
of less than 30$^\circ$ around the forward and backward direction.
With Cohen's distribution the shape is too flat, whereas the Wigner
and the product distribution overestimate the forward direction. Both
electrons have the same small final energy, which means that they
influence each other for a long time. Consequently, these cross
sections are more dominated by the post collisional interaction (PCI)
than those at higher energies. In order to describe PCI effects
correctly, the relative energy of the outgoing electrons is important.
Due to our construction of the cross section in terms of the energy
transfer this ratio is correctly described only for initial
distributions on the energy shell (i.e. without an energy spread).

\subsubsection{Triple differential ionization cross section}

At this impact energy experiments exist for a variety of geometries
\cite{ROD96,ROD93}. Table \ref{tbl:TDCSParams17} summarizes them for
the cross sections shown. For an overview over recent quantum
calculations see \cite{BAE01b, BRA00a}.

\begin{table}[h]
	\centering
	\begin{tabular}{c|cl}
		geometry & extraction parameters\\
		\hline \\[-0.2cm]
		symmetric & $|\theta_a - \theta_b|$ $\leq$ 8$^\circ$,
			$|E_a-E_b|$ $\leq$ 0.3 eV \\[0.2cm]
		asymmetric & $\theta_a$ = 60$^\circ$ $\pm$ 3$^\circ$,
			$E_b$ = 2 eV $\pm$ 0.2 eV \\[0.2cm]
		const. $\theta_{ab}$ & $\theta_{ab}$ = 90$^\circ$, 180$^\circ$
			$\pm$ 5$^\circ$, $E_b$ = 2 eV $\pm$ 0.2 eV
	\end{tabular}
	\caption{Overview over the shown geometries and parameters for the
	TDCS at $E_{in}$ = 17.6 eV. For all cross sections the scattering
	plane is confined to $\pm$ 12$^\circ$. The measurements were done
	by R\"oder etal \cite{ROD96, ROD93}}
	\label{tbl:TDCSParams17}
\end{table}
\showlabel{tbl:TDCSParams17}

\epsfxsize=13cm
\epsfysize=14.8cm
\begin{figure}[htb]
	\centering
	\hspace{0.5cm} \epsffile{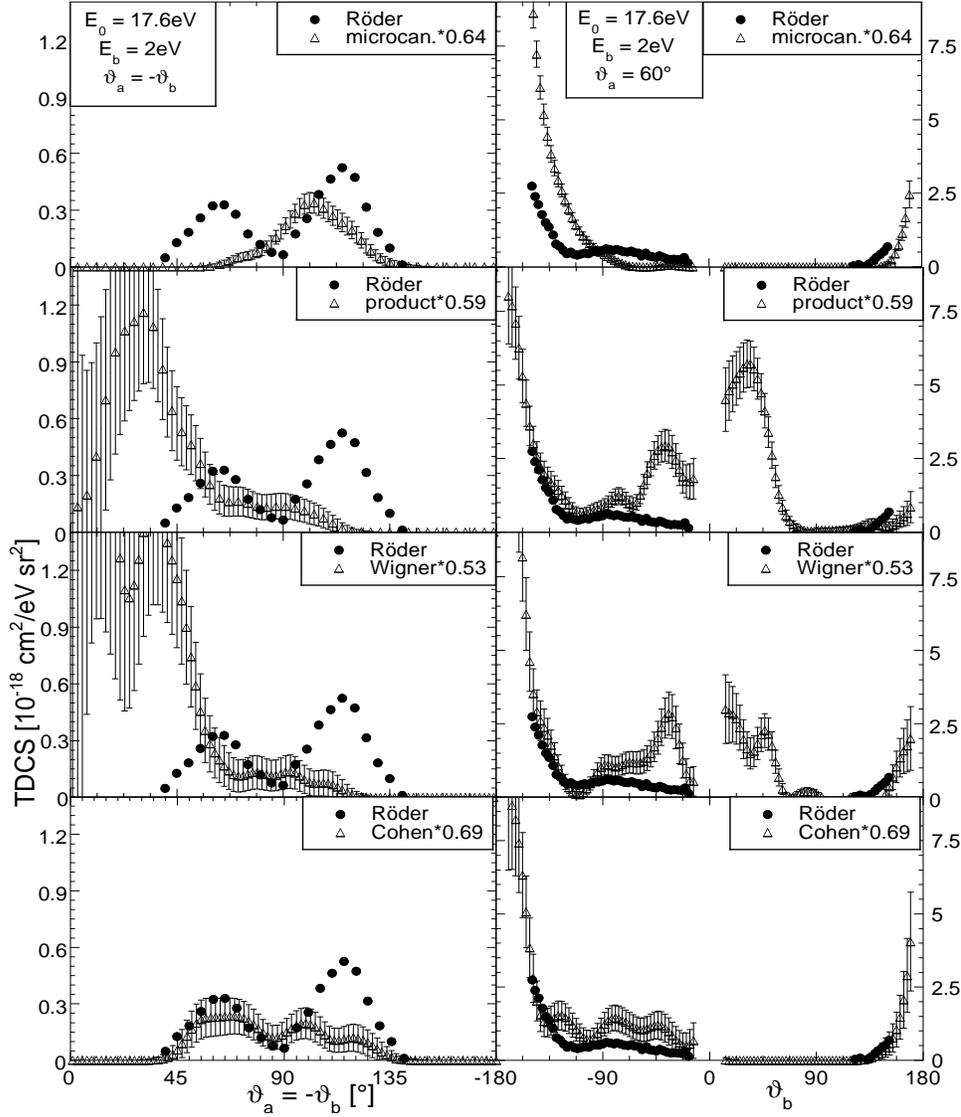}
	\caption{TDCS for a total energy of 4 eV: fully symmetric, i.e.
	$\theta_a$ = $\theta_b$ and $E_a$ = $E_b$ = 2 eV, on the left side
	and on the right side in Ehrhardt geometry for $E_b$ = 2 eV and
	$\theta_a$ = 60$^\circ$ for the four initial distributions. The
	cross sections are scaled onto the total cross section.}
	\label{fig:TDCS17-s60}
\end{figure}
\showlabel{fig:TDCS17-s60}

\epsfxsize=13cm
\epsfysize=14.8cm
\begin{figure}[htb]
	\centering
	\hspace{0.5cm} \epsffile{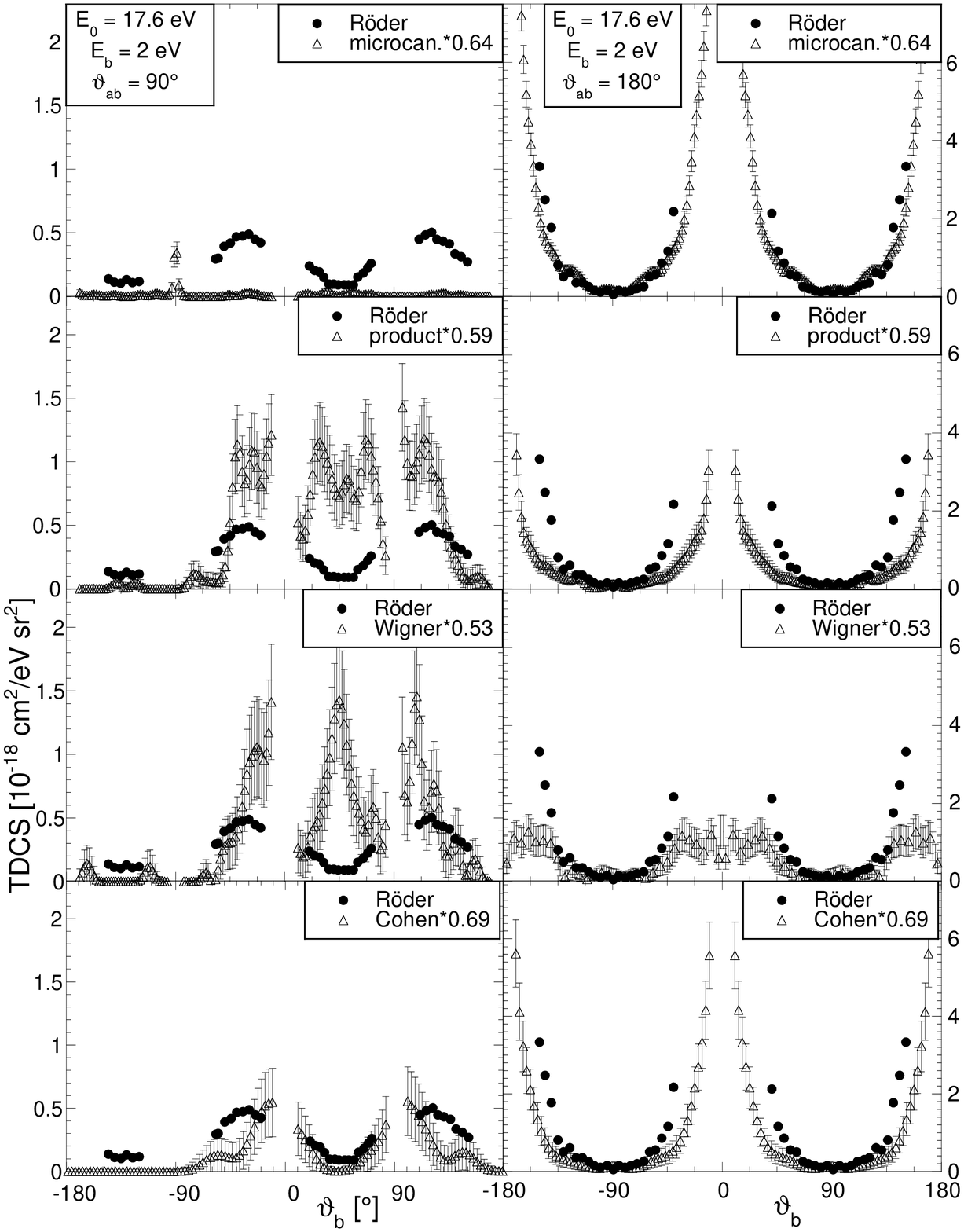}
	\caption{TDCS for constant angle between the electrons, left with
	$\theta_{ab}$=90$^\circ$, right at 180$^\circ$. The cross sections
	are scaled onto the total cross section.}
	\label{fig:TDCS17-tab}
\end{figure}
\showlabel{fig:TDCS17-tab}

For the Wigner and the product distribution the same behavior as at
$E_{in}$ = 54.4 eV can be observed, but now the fraction of electrons
going into the same direction is much larger, see the fully symmetric
cross section and in Ehrhardt geometry, \fig{fig:TDCS17-s60}. This
should not occur due to the mutual repulsion of the electrons.
However, for those initial distributions which are not restricted to
the energy shell (all but the microcanonical one) the energy of the
initially bound electron calculated from the energy transfer does not
necessarily correspond to the actual momentum of the electron. Hence,
the two electrons may leave the nucleus one after the other, although
they have the same ``nominal'' energy. This problem does not occur for
the microcanonical distribution which is on the energy shell.

At 17.6 eV impact energy none of the initial distributions can fully
reproduce the cross sections, but in the overall picture Cohen's
distribution plays best: the energy spread is on the one hand small
enough to prevent the electrons from going into the same direction and
on the other hand the phase space density is still accurate enough for
the ``fast'' binary ionization process. Hence, Cohen's distribution is
the only one which predicts the location and widths of binary and
recoil peaks although not their height (\fig{fig:TDCS17-s60}). At
constant $\theta_{ab}$ for all but Cohen's distribution the (relative)
magnitude of the cross section for different angles is grossly wrong
(\fig{fig:TDCS17-tab}). For the Wigner and the product distribution
even the shape of the cross section is in disagreement with the
experiment at $\theta_{ab}$ = 90$^\circ$.

The cross sections at 17.6 eV show that even at this low total energy
the fully differential cross section is sensitive to the initial phase
space density. On the other hand Wannier's arguments suggest that
close to threshold the cross sections depend primarily on the final
phase space configuration determined by the long time behavior of the
ionization dynamics \cite{WAN53}. To describe it correctly the
accurate energy of the outgoing electrons is crucial. Hence, the
initial distribution must have a minimal energy spread. At 4 eV above
threshold these two (contradicting) requirements are fulfilled best by
Cohen's energy distribution.


\section{Summary and outlook}

We have derived and tested a quasiclassical ansatz for charged
particle impact ionization. Starting from the time-dependent quantum
mechanical formulation in the quantum Wigner picture we can reproduce
in the limit $\hbar=0$ the standard Classical Trajectory Monte Carlo
(CTMC) method. However, our ansatz is more general. By keeping the
backward--forward propagation scheme of the M\o{}ller formalism and by
defining the cross section in terms of energy transfer we can exploit
the full classical limit ($\hbar=0$) of the Wigner distribution for
the initial state, which is neither stationary in the classical
approximation nor confined to the energy shell, two properties which
are necessary for standard CTMC. In fact, our formulation allows one
to use arbitrary initial phase space distributions. We have calculated
and compared fully differential classical ionization cross sections
for four initial state distributions: the quantum Wigner function, the
(standard) classical microcanonical distribution and two other
distributions which interpolate to some extent between Wigner and
microcanonical distribution. From our results we conclude that at high
impact energies the density in phase space is the crucial property to
describe the ionization process. For lower energies the correlation
between coordinate and momentum becomes increasingly important. Even
at the low energy of 4 eV above threshold the initial state has to be
modeled correctly in coordinate and momentum space, although the long
time evolution of the slow ionized electrons gains more and more
influence on the dynamics.

The classical calculations are not meant to be a high quality
alternative for quantum treatments; rather we wanted to formulate and
test a consistent quasiclassical scattering theory, applicable to the
most differential cross sections, and free from the limitation to one
active target electron of previous approaches. Yet, most of the
technical expertise from previous CTMC calculations can be used with
only minor modifications in the description of the initial target
state, the propagation scheme and the extraction of the cross
sections.

With the present work we have demonstrated for one-electron targets,
namely hydrogen, that the aforementioned goals can be achieved.
However, the main advantage of being able to deal with unstable
initial distributions will become apparent when calculating double
ionization of two-electron atoms. The classical autoionization of the
bound electrons does not affect our quasiclassical approach. What
should be improved in future work, particularly if one is interested
in collisions with small excess energy, is the implementation of
contributions off the energy shell. Since we have seen that less
differential cross sections tend to be better approximated with the
quasiclassical approach we are optimistic that a quasiclassical triple
differential cross section will be more accurate for double ionization
than for single ionization. However, this remains to be proven in
future work.

\smallskip

Most of this work was funded by the Deutsche Forschungsgemeinschaft
through the SFB 276 at the University of Freiburg, Germany.

\bigskip


\end{document}